# Spontaneous emission dynamics of Eu$^{3+}$ ions coupled to hyperbolic metamaterials

Gabriel I. López-Morales,[1,2,3] Mingxing Li,[1] Ravindra K. Yadav,[4] Harshavardhan R. Kalluru,[4]
Jaydeep K. Basu,[4] Carlos A. Meriles,[1,3] Vinod M. Menon[*,1,3]

[1]Department of Physics, City College of the City University of New York, New York, NY 10031 USA
[2]Department of Chemistry, Lehman College of the City University of New York, Bronx, NY 10468, USA
[3]The Graduate Center of the City University of New York, New York, NY 10016, USA
[4]Department of Physics, Indian Institute of Science, Bangalore 560012, India
[*]email: vmenon@ccny.edu

**ABSTRACT**: Sub-wavelength nanostructured systems with tunable electromagnetic properties, such as hyperbolic metamaterials (HMMs), provide a useful platform to tailor spontaneous emission processes. Here, we investigate a system comprising Eu$^{3+}$(NO$_3$)$_3$·6H$_2$O nanocrystals on an HMM structure featuring a hexagonal array of Ag-nanowires in a porous Al$_2$O$_3$ matrix. The HMM-coupled Eu$^{3+}$ ions exhibit up to a 2.4-fold increase of their decay rate, accompanied by an enhancement of the emission rate of the $^5D_0 \rightarrow ^7F_2$ transition. Using finite-difference time-domain modeling, we corroborate these observations with the increase in the photonic density of states seen by the Eu$^{3+}$ ions in the proximity of the HMM. Our results indicate HMMs can serve as a valuable tool to control the emission from weak transitions, and hence hint at a route towards more practical applications of rare-earth ions in nanoscale optoelectronics and quantum devices.

Thanks to the screening of their partially filled 4$f$ orbitals, rare-earth (RE) ions are found to exhibit atom-like optical transitions, often insensitive to the choice of the host matrix. This behavior is best illustrated in crystals such as Y$_2$SiO$_5$, where Eu$^{3+}$ ions are seen to feature optical linewidths as narrow as ~400 Hz in the visible region of the spectrum.[1] Because 4$f$ electrons are often unpaired, they also exhibit large magnetic moments that can be typically polarized and manipulated with high fidelities. These properties have prompted interest for applications in quantum information processing,[2–5] particularly as quantum memories, with nuclear-spin-based coherent storage times exceeding several hours.[6]

A main consequence of their narrow optical linewidths, however, is that the fluorescence from RE emitters is intrinsically faint, hence complicating the isolation of single-ions and their spin manipulation.[7–9] In an attempt to overcome these limitations, recent work has demonstrated enhanced detection via coupling to photonic-crystal cavities based on Purcell enhancement of the spontaneous emission.[9–11] Nonetheless, the fixed resonance modes dictated by the cavity structure, and the complexity of the required nanofabrication techniques hinder the practicality of this approach.

Alternatively, hyperbolic metamaterials (HMM), usually constructed by sub-wavelength stacking of metal and dielectric interfaces, result in a broadband divergence of the photonic density of states (PDOS).[12–14] This phenomenon stems from their hyperbolic dispersion in momentum space, supporting the propagation of high-$k$ electromagnetic waves that are otherwise evanescent.[13,15] Additionally, this characteristic broadband response in HMMs can be conveniently tuned based on the choice of the metal/dielectric interface and the corresponding metal fill-factor, all of which can be leveraged so as to gain control on the spontaneous emission of multi-level systems.[14–16]

Here, we investigate the emission dynamics of a set of Eu$^{3+}$(NO$_3$)$_3$·6H$_2$O nanoparticles overlaid on an HMM structure formed by a hexagonal array of Ag-nanowires grown within a porous Al$_2$O$_3$ matrix.[17] Compared to particles on a bare SiO$_2$ substrate, we observe a 2.4-fold increase of the Eu$^{3+}$ spontaneous emission rate near the HMM, accompanied by an enhancement of the fluorescence from the $^5D_0 \rightarrow ^7F_2$ Eu$^{3+}$ transition. With the help of finite-difference time-domain (FDTD) simulations, we show this response is consistent with that expected for HMM-coupled Eu$^{3+}$ ions. Further, our calculations suggest there is much room for improvement provided the HMM geometry is revised to match the emission properties of $Eu^{3+}$ more closely. As a whole, these observations highlight intriguing opportunities on the use of HMMs to tailor the emission properties of rare-earth ions for applications in quantum information processing and nanoscale metrology.

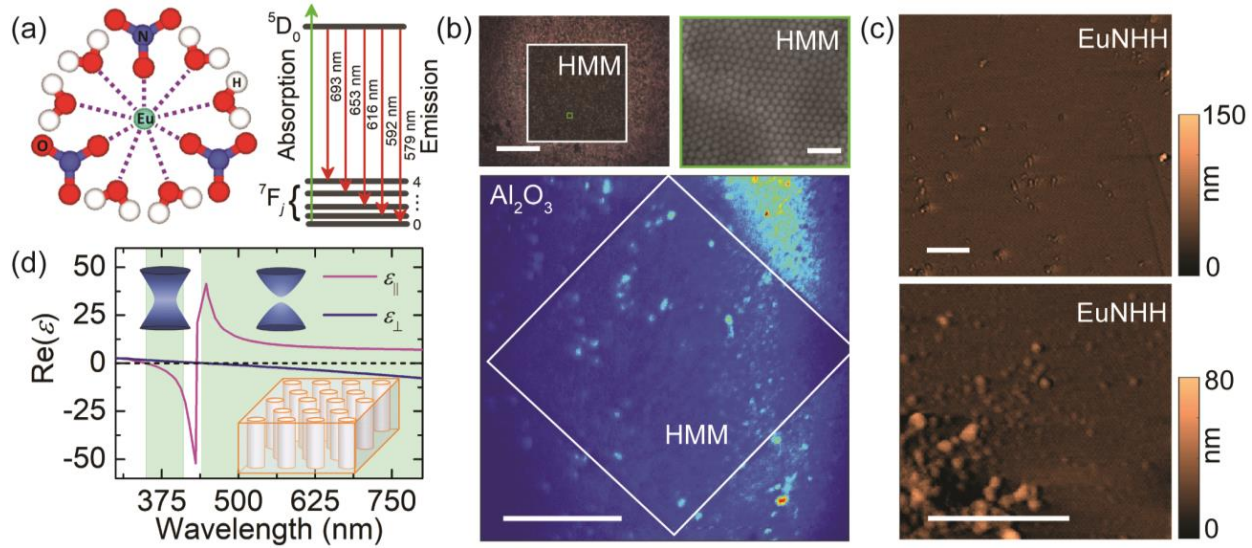

**Figure 1. $Eu^{3+}$ ions on an HMM substrate.** (a) Schematic representation of $Eu^{3+}(NO_3)_3 \cdot 6H_2O$ (ENHH) alongside a simplified energy-level diagram for $Eu^{3+}$ under 532nm-excitation. Crystal-field interactions cause the $^7F$ level to split into its $j$-components (only $j$ = 0–4 sublevels are included). (b) Optical, scanning-electron, and confocal microscopy (upper left, upper right, and bottom images, respectively) of the nanowire HMM used in this work. White squares highlight the active HMM region. (c) Atomic-force microscopy (AFM) of ENHH particles. Scale bars in optical/confocal, scanning-electron microscopy, and AFM images correspond to 40, 5 and 0.5 μm, respectively. (d) Real part of the parallel (pink) and perpendicular (blue) components of the permittivity tensor for a 75-nm Ag -nanowire/$Al_2O_3$ HMM (inset structure). In our case, an estimated inter-wire distance of 125 nm results in a fill-factor of 0.32, supporting HMM modes (green-shaded regions) within 350–405 nm (type-II HMM) and above 500 nm (type-I HMM). Within these regions, the anisotropic nature of the permittivity tensor results in hyperbolic iso-frequency surfaces in momentum space (inset hyperbolas).

We focus on the luminescence properties of $Eu^{3+}$ ions in crystals of $Eu^{3+}(NO_3)_3 \cdot 6H_2O$ (europium nitrate hexa-hydrate or ENHH for short). A schematic of the atomic structure is shown in Fig. 1(a), along with a simplified version of the energy-level diagram for $Eu^{3+}$ under 532 nm-excitation. In this case, crystal-field interactions cause the $Eu^{3+}$ $^7F$ level to split into its $j$ = 0–6 components. 532 nm-illumination (green arrow) results in excitation above the $^5D_0$ level, which is mostly followed by a non-radiative decay into $^5D_0$ followed by emission from one of the $^5D_0 \rightarrow {^7F_j}$ radiative transitions (red arrows).[18–20] In some cases, direct radiative decay from the $^5D_1$ is non-negligible and can result, for example, in overlap with some of the $^5D_0 \rightarrow {^7F_j}$ transitions.[21,22] The $^5D_1 \rightarrow {^7F_j}$ transitions, however, are usually much faster than those stemming from $^5D_0$ and can thus be identified via time-resolved (TR) measurements.[20,21]

Our glass substrate includes a region of porous alumina ($Al_2O_3$) inside which a hexagonal array of 75-nm silver (Ag) diameter nanowires is grown to form the HMM (white square in Fig. 1(b) upper-left panel). A scanning electron microscopy (SEM) image of a small section of the HMM is shown in Fig. 1(b) (green square, upper-right panel). From the SEM scan we estimate an inter-wire distance of 125 nm, resulting in a fill-factor of 0.32.

To bring the RE ions in close contact with the HMM structure, we start by atomizing a low-concentration solution (0.018 M) of ENHH (Sigma Aldrich) in ethanol onto the HMM substrate to create particles of variable size. Rapid evaporation of the solvent produces a collection of crystals distributed throughout the substrate. The lower panel in Fig. 1(b) shows a fluorescence image of an area that includes the HMM section of the substrate: Selection of ethanol as the solvent proved critical to deposit enough droplets on the HMM, suggesting electrostatic charging (see below). AFM imaging of the ENHH particles (Fig. 1(c)) reveals that while some nanoparticles are isolated and uniformly distributed, others group to form flat particle aggregates or poly-crystalline films, with an average height between 30–50 nm.

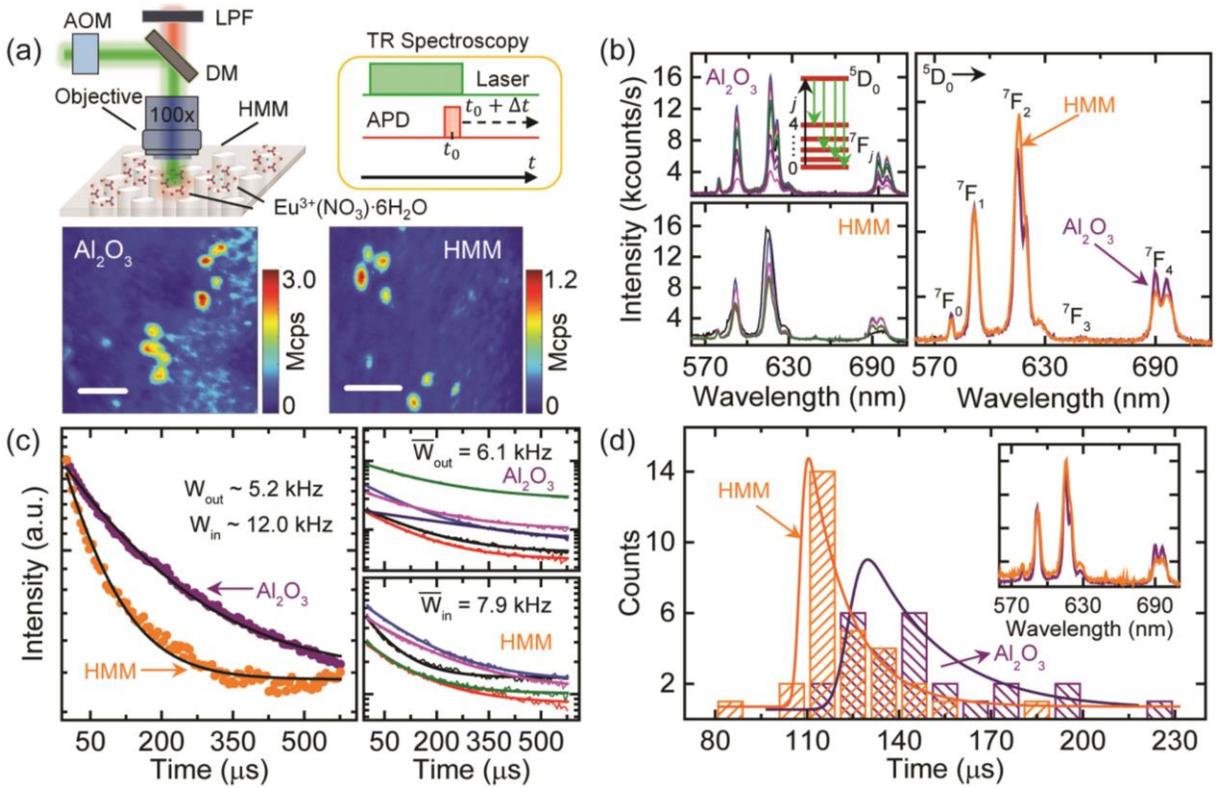

**Figure 2. Controlling photoluminescence from Eu$^{3+}$ ions using an HMM.** (a) (Top) Experimental setup (left) and TR spectroscopy protocol (right). (Bottom) Representative fluorescence images from areas within (right) and outside (left) the HMM region. Scale bars represent 5 µm. (b) Emission spectra from individual ENHH clusters outside (upper left-side panel) and inside (lower left-side panel) the HMM region. A comparison between representative spectra outside and inside the HMM (purple and orange solid traces, respectively) is shown in the right-side panel of (b). (c) (Left panel) TR spectroscopy from select clusters in (a) illustrating two extreme cases in and out of the HMM (purple and orange traces, respectively). (Upper/lower right-side panels) Representative collection of fluorescence decay curves. Solid lines are fits to single-exponential decays. (d) Histogram representation of the of excited state lifetime distribution; solid lines are guides to the eye. Inset: spectral signatures corresponding to the decay curves of the two extreme cases in and out of the HMM (purple and orange traces, respectively) shown in left panel of (c).

An important aspect of this study is to ensure that the fluorescence from Eu$^{3+}$ ions lies within the hyperbolic dispersion regime of the HMM. To characterize this, we employ effective medium theory.[23] Fig. 1(d) shows the real part of the parallel (magenta) and perpendicular (blue) components of the permittivity tensor for a 75-nm Ag-nanowire/Al$_2$O$_3$ HMM with a fill-factor of 0.32 (inset structure). We find that this structure supports HMM modes (green-shaded regions) within 350–405 nm (type-II HMM) and above 500 nm (type-I HMM). Within these regions, anisotropies of the effective permittivity tensor result in hyperbolic iso-frequency surfaces in momentum space (inset hyperbolas).[12]

In our experiments, we use a 100x, 1.30NA oil-immersion objective (Nikon, Plan Fluor) to excite the Eu$^{3+}$ ensembles via a 532-nm continuous-wave (CW) laser (Fig. 2(a)). The collected fluorescence is then filtered out via a dichroic mirror (DM) combined with a long-pass filter (LPF). We carry out time-resolved (TR) spectroscopy measurements using an acousto-optic modulator (AOM) to time the laser excitation, and a pulse controller to gate the avalanche photo-detector (APD, Excelitas) over a pre-defined, short time window $\Delta t$ (20 µs). Fluorescence decay curves emerge from gradually varying the delay of the acquisition interval $\Delta t$ relative to the end of the excitation laser pulse.

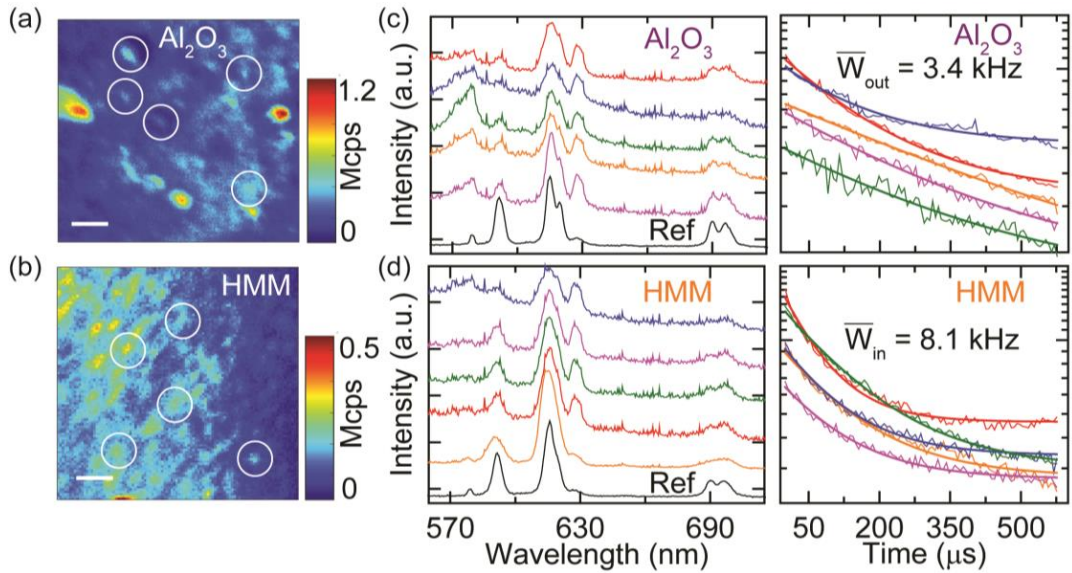

**Figure 3. Particle size effects.** (a, b) Confocal fluorescence images of small (< 1 μm lateral size) nanocrystals outside (a) or inside (b) the HMM area. Scale bars are 5 μm. (c) Optical spectra (left) and fluorescence decay curves (right) from the particles circled in (a). A reference spectrum (Ref, bottom trace in the left panel) has been included for comparison; the average decay rate reaches 3.4 kHz. (d) Same as in (c) but for the particles circled in (b). Here the average time constant amounts to 8.1 kHz, shorter than in Fig. 2. Both in (c) and (d) we find substantial broadening in the emission lines, along with varying peak-intensity ratios between transitions.

Confocal fluorescence images reveal strong luminescence from individual ENHH clusters both outside and inside the HMM area (lower left-/right-side panels in Fig. 2(a)). Optical spectroscopy from these sites indicates $Eu^{3+}$ emission from the $^5D_0 \rightarrow ^7F_j$ radiative transitions (Fig. 2(b) inset). A comparison between the emission spectra from particles outside or inside the HMM does not expose major differences, though the latter tend to show broader features (likely deriving from heterogeneous shifts in the emission wavelengths, see below). On the other hand, fluorescence lifetime measurements over a similarly large particle set show that, on average, the decay time of $Eu^{3+}$ ions is shorter within the HMM region. This is shown in Fig. 2(c) where we observe differences between the characteristic decay rates by up to a factor 1.3. The histogram in Fig. 2(d) captures these changes more quantitatively: In both cases we observe broad, skewed distributions of time constants though the maximum is shifted to lower values for particles sitting within the HMM structure. Overall, this shift corresponds to an excited state lifetime reduction of about 20%.

Interestingly, the above findings do not necessarily correlate with changes in fluorescence intensity, which, as shown in Fig. 2(b), responds differently depending on the emission wavelength. For example, while the emission peak at 620 nm ($^5D_0 \rightarrow ^7F_2$ transition) experiences a growth of ~20%, virtually no change is observed at shorter wavelengths (transitions $^5D_0 \rightarrow ^7F_0$ and $^5D_0 \rightarrow ^7F_1$); further, despite the shorter lifetimes (observed even for spectrally selective measurements, not shown for brevity), the emission brightness decreases for the $^5D_0 \rightarrow ^7F_4$ transition at longer wavelengths (see emission peaks at 695 nm).

To characterize possible size-effects on the overall luminescence properties, we now turn our attention to small-radius ENHH particles (< 1 μm lateral size). Emission spectra from these nanoparticles (left-side panel in Fig. 3(c)) circled in Fig. 3(a), and (left-side panel in Fig. 3(d)) circled in Fig. 3(b) reveal marked differences on the luminescence properties of $Eu^{3+}$ ions. For example, direct comparison between each spectrum set and a spectrum representative of larger-radius particles (>2 μm lateral size), identified as "Ref" in Figs. 3(c, d)) reveals substantial broadening of identified transitions, both inside and outside of the HMM region, along with varying peak-intensity ratios.

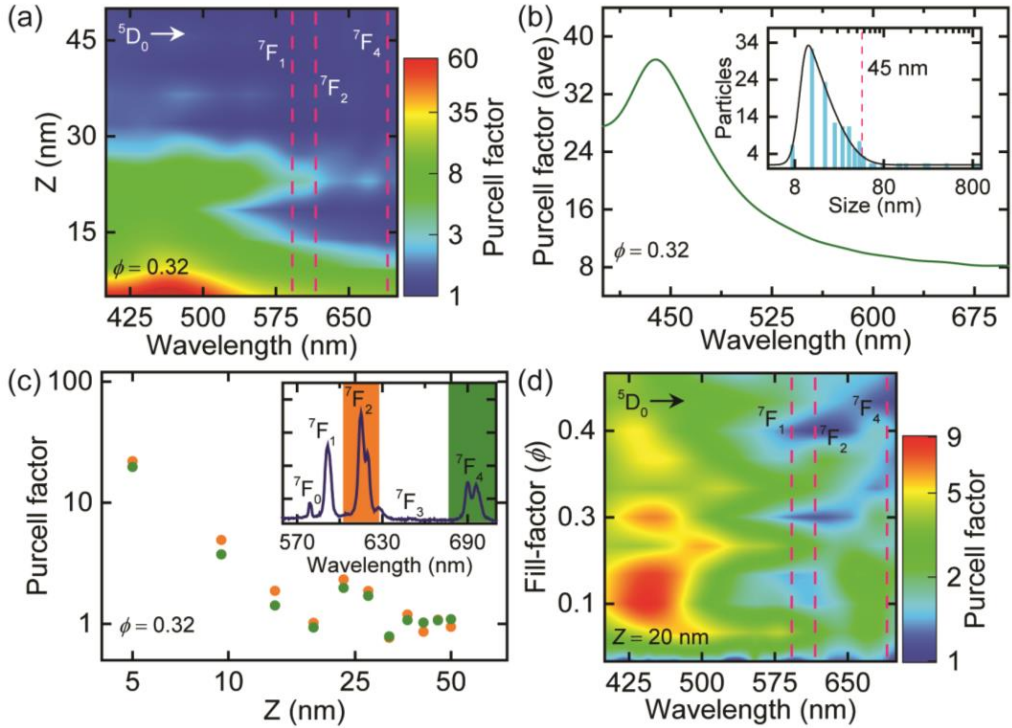

**Figure 4. Calculated Purcell enhancement.** (a) Simulation of the Purcell factor for a dipole placed parallel on top of the Ag-nanowire/$Al_2O_3$ HMM structure (fill-factor of 0.32). (b) Calculated Purcell factor for a dipole placed on the HMM surface, averaged over the height distribution shown in the inset. Inset: particle-size distribution from AFM imaging of ENHH particles (see Fig. 1(c)). (c) Purcell factor for dipole emission from the $^5D_0 \rightarrow {}^7F_2$ or $^5D_0 \rightarrow {}^7F_4$ transitions in $Eu^{3+}$ as a function of the distance Z from the HMM surface. The color bands in the inset indicate the wavelength range used in each case. (d) Purcell factor for a lateral dipole placed 20-nm from the HMM surface as a function of the fill-factor and wavelength.

As expected, we also find a size-related effect on the effective Purcell enhancement of the $Eu^{3+}$ ions. For instance, in the case of small-radius particles inside the HMM, we find an average enhancement of 2.5 times in the emission rate on the HMM (8.1 kHz) compared to the outside (3.4 kHz on average). In contrast, for larger-radius particles, the enhancement in Purcell factor is negligible (~1.3). These observations are consistent with previous measurements[24] and FDTD simulations, and stem from the reduction of averaged contributions over ions located farther away from the active HMM surface (where the effective Purcell factor is ~1).

To better understand the interplay between $Eu^{3+}$ emission and the HMM structure, we use the FDTD method to calculate the Purcell factor for a point dipole as a function of the distance Z to the HMM surface, the emission wavelength λ, and the HMM's fill-factor ϕ. The presence of closely spaced Ag-nanowires within the $Al_2O_3$ matrix creates a strong birefringence and results in a hyperbolic dispersion that profoundly changes the PDOS both inside the HMM and near its surface.[13–15,17,25] Placing an emitter in close proximity to the HMM thus leads to a broadband enhancement of spontaneous emission and a corresponding increase of the decay rate through the Purcell effect.[24,26,27]

Fig. 4(a) shows a 2D color map of the calculated Purcell factor for a point dipole parallel to the HMM surface as a function of the distance and wavelength. We first consider the case ϕ = 0.32, close to the experimental value. Near the HMM, we find the broadband response anticipated for hyperbolic media. Our calculations indicate maximum Purcell enhancement within the 400–500 nm window, even though the HMM impact extends to the range of longer wavelengths overlapping with the $Eu^{3+}$ emission spectrum. Naturally, the enhancement decreases quickly with increasing distance, meaning that the observed changes in a given particle derive from the fraction of emitters in close contact with the HMM. Fig. 4(b) shows an average Purcell factor of dipoles placed above the HMM surface based on the particle-size distribution as extracted from AFM imaging (inset of Fig. 4(b)). For $Z < 5$ nm, the average Purcell factor shows enhancements of 37 and 8 around 430 nm and 640 nm, respectively, with Purcell factor of ~8 in a wide wavelength range due to the broadband divergence of the PDOS in HMMs.

The experimental results summarized in Fig. 2(b) indicate an increase (decrease) in emission intensities for the 620 nm (690 nm) transitions. To explain these observations, we extract the Purcell factor as a function of distance from the HMM surface for transitions centered at 620 nm and 690 nm (two of those identified in Fig. 4(a)). As shown in Fig. 4(c), the Purcell factor is slightly higher at 620 nm than at 690 nm, indicating a stronger enhancement. Therefore, for our HMM structure, we expect an asymmetric effect on the $^5D_0 \rightarrow {}^7F_2$ and $^5D_0 \rightarrow {}^7F_4$ transitions in $Eu^{3+}$, with a net increase on the photon-count for the former transition, consistent with our observations.

Better adapting the HMM structure to the emission range of $Eu^{3+}$ could potentially increase the effective fluorescence enhancement in the visible range.[14] For instance, Fig. 4(d) describes the Purcell factor for a dipole located 20 nm from the HMM surface, as a function of $\phi$ and emission wavelength. This distance of 20 nm corresponds to half the average height (dipole central position) obtained from the ENHH particle-size distribution in Fig. 4(b). We find that by tuning the fill-factor between 0.08 and 0.25 one could further increase the effective Purcell factor for the wavelength range that accommodates all the $^5D_0 \rightarrow {}^7F_j$ transitions in $Eu^{3+}$.

In summary, we investigated the effects of HMM on the luminescence properties of $Eu^{3+}$ ions in $Eu^{3+}(NO_3)_3 \cdot 6H_2O$ particles. Upon 532-nm excitation, we observe strong photoluminescence with well-defined $^5D_0 \rightarrow {}^7F_j$ ($j$ = 0–4) transitions at room-temperature. For small-radius particles (< 1 μm), where the effect of the HMM is more pronounced, we observed 2.4 times enhancement in the spontaneous emission rate. From FDTD simulations, we find that the net decay-rate enhancement of $Eu^{3+}$ ions could be substantially increased by tuning $\phi$ between 0.05–0.3 for the HMM structure.

**Acknowledgements**: M.L. and V.M.M. acknowledge support from the National Science Foundation through grant NSF-1906096; C.A.M acknowledges support from the Department of Energy through grant BES-DE-SC0020638. All authors acknowledge access to the facilities of the NSF CREST IDEALS, grant number NSF-HRD-1547830. V.M.M. and J.K.B. also acknowledge the program Scheme for Promotion of Academic and Research Collaboration (SPARC), Ministry of Human Resources Development (MHRD), India, for supporting travel between India and the US to foster international collaboration. We thank Daniela Pagliero and Artur Lozovoi for assistance with the experimental setup. We are also indebted to Fernando Meneses for useful discussions on the HMM structure.

**Data availability**: The data that support the findings of this study are available from the corresponding author upon reasonable request.